\documentstyle[preprint,aps,prb]{revtex}

\begin{document}
\draft
\preprint{}
\title{
\vspace{-.5in}
    A computational study of 13-atom Ne-Ar cluster \\
    heat capacities}
\author{D. D. Frantz}
\address{Department of Chemistry, University of Waterloo, \\
    Waterloo, Ontario N2L 3G1, Canada}
\date{\today}

\maketitle

\vspace{-.25in}
\begin{abstract}
Heat capacity curves as functions of temperature were calculated using Monte
Carlo methods for the series of Ne$_{13-n}$Ar$_n$ clusters ($0 \leq n \leq
13$). The clusters were modeled classically using pairwise additive
Lennard-Jones potentials. The J-walking (or jump-walking) method was used to
overcome systematic errors due to quasiergodicity. Substantial discrepancies
between the J-walking results and those obtained using standard Metropolis
methods were found. Results obtained using the atom-exchange method, another
Monte Carlo variant for multi-component systems, also did not compare well
with the J-walker results. Quench studies were done to investigate the
clusters' potential energy surfaces. Only those Ne-Ar clusters consisting
predominately of either one or the other component had lowest energy isomers
having the icosahedral-like symmetry typical of homogeneous 13-atom rare gas
clusters; non-icosahedral structures dominated the lowest-energy isomers for
the other clusters. This resulted in heat capacity curves that were very much
different than that of their homogeneous counterpart. Evidence for
coexistence behavior different than that seen in homogenous clusters is also
presented.
\end{abstract}

\narrowtext

% Ar-Kr Cv paper: introduction section
\section{Introduction}
The investigations discussed in this report are the second part of a
comprehensive study of 13-atom binary rare gas clusters; the first part
examined the series of clusters Ar$_{13-n}$Kr$_n$ for $0 \leq n \leq
13$.\cite{ArKr_Cv} Thirteen-atom
homogeneous\cite{Magic_Cv,coexist,DJB,Tsai-Jordan,J-walker,NP,QS,EK,BB}
and
heterogeneous\cite{LMA,FF,GPC,GHJ,TZKDS,TAP,Lopez-Freeman,LeRoy,SLA,CSLR}
clusters have been the subject of many studies. The homogeneous rare gas
clusters exhibit ``magic number''
effects\cite{Magic_Cv,CSLR,magic_coexist,HA,Northby,FD_magic,ESR} for many of
their properties that are due in large part to their compact icosahedral
ground state configuration. Because Ar and Kr have similar sizes (the Ar
radius is about 11\% smaller than the Kr radius), all the 13-atom Ar-Kr
clusters also had icosahedral-like lowest-energy configurations that played
dominant roles in their behavior. This size similarity resulted in two
different categories of Ar-Kr isomers: topological isomers based on geometric
structures that were similar to those of their homogeneous counterparts, and
permutational isomers, which were based on the various rearrangements of the
different component atoms within a topological form. Ne and Ar have quite
different sizes (the Ne radius is about 19\% smaller than the Ar radius).
This results in many additional non-icosahedral isomers with energies similar
to the icosahedral-like isomers, which dramatically alters the Ne-Ar cluster
behavior compared to that of their Ar-Kr counterparts. The other major
difference between Ne-Ar and Ar-Kr clusters is in their intermolecular
potentials, where the Ne and Ar values are far more dissimilar compared to
the Ar and Kr values ($\epsilon_{\mbox{\scriptsize Ne-Ne}} /
\epsilon_{\mbox{\scriptsize Ar-Ar}} = 0.2982$, $\epsilon_{\mbox{\scriptsize
Ar-Ar}} / \epsilon_{\mbox{\scriptsize Kr-Kr}} = 0.7280$).

Heat capacities have been useful for elucidating the nature of cluster
solid-liquid ``phase''
transitions.\cite{Magic_Cv,DJB,Tsai-Jordan,J-walker,NP,QS,Lopez-Freeman,Matro_Freeman,Lopez,JWFPI}
Heat capacity curves as functions of temperature for homogeneous
13-atom clusters are characterized by a very large peak in the
solid-liquid transition region that is primarily a consequence of the large
energy gap between the ground state icosahedral isomer and the higher energy
non-icosahedral isomers.\cite{Magic_Cv,DJB,Tsai-Jordan,J-walker,NP,QS}
Thirteen-atom Ar-Kr clusters have heat capacity curves that are mostly similar
to the homogeneous cluster curve, but there are interesting differences
arising from the effects of the permutational isomers, made manifest as
an additional, very small, heat capacity peak occurring at very low
temperatures.\cite{ArKr_Cv,Lopez-Freeman} This feature, arising from the
low-energy icosahedral-like permutational isomers, is reminiscent of
order-disorder transitions known to occur in some bulk alloy
materials.\cite{Lopez-Freeman} For 13-atom Ne-Ar clusters, though, the large
disparity in the Ne and Ar potential parameters results in qualitatively
different heat capacity curves.

Cluster heat capacities are notoriously difficult to calculate accurately
using simulation methods because of poor convergence due to the incomplete
sampling of configuration space. This is especially so for heterogeneous
clusters, where the inadequate mixing of the different component atoms can
become problematic at lower temperatures, leading to large systematic errors.
As with my previous Ar-Kr heat capacity studies, I used the J-walking (or
jump-walking) method\cite{J-walker} to obtain the heat capacities to good
accuracy. J-walking has been shown to greatly increase Monte Carlo
convergence rates and overcome systematic errors due to incomplete sampling
in several
studies.\cite{ArKr_Cv,Magic_Cv,Tsai-Jordan,Lopez-Freeman,Matro_Freeman,Lopez,JWFPI,SLF}
Despite the poor performance of the standard Metropolis method in calculating
heat capacities, the results obtained from the method can still provide
additional insights when compared with the J-walking results, and so I also
calculated heat capacity curves for each Ne-Ar cluster using the standard
Metropolis method, as well as with a modified version that augments binary
cluster simulations by using atom-exchange methods to enhance the mixing of
the components.\cite{TAP}

I begin in Section~\ref{Sec:theory} with a brief overview of the various
computational methods I used. Section~\ref{Sec:results} first examines the
structural properties of 13-atom Ne-Ar clusters, discusses the results
obtained for their heat capacities and potential energies as functions of
temperature, and then examines the results obtained from the quench studies.
The results are also compared to their corresponding Ar-Kr results.
Finally, in Section~\ref{Sec:conclusion}, I summarize my findings and discuss
some of the insights provided by the Ne-Ar cluster results.

% Ne-Ar Cv paper: theory section
\section{Computational Method}                        \label{Sec:theory}
The computational methods used in this work were mostly similar to the
methods I used in my previous study of Ar-Kr clusters.\cite{ArKr_Cv} The
following briefly summarizes the important points; the reader is referred to
the previous work for a more detailed description.

Monte Carlo simulations were run for all clusters in the series
Ne$_{13-n}$Ar$_n$ with $0 \leq n \leq 13$. The clusters were modeled by the
usual pairwise additive Lennard-Jones potential. Table~\ref{Tbl:LJ-params}
lists the Lennard-Jones parameters.\cite{LDW} The Ne-Ar interaction
parameters were determined from the usual Lorentz-Berthelot mixing
rules.\cite{Allen-Tild}. I have followed Lee, Barker and Abraham\cite{LBA}
and have confined the clusters by a perfectly reflecting constraining
potential of radius $R_c$ centered on the cluster's center of mass. To
maintain a common set of boundary conditions throughout the survey, the
constraining radius was identical for all the clusters studied, with $R_c =
3\sigma_{\mbox{\scriptsize\rm Ar-Ar}}$. Analogous to the Ar-Kr study, the
energies and heat capacities are expressed in reduced units in terms of the
smaller atom interaction, with $V^* = V/\epsilon_{\mbox{\scriptsize Ne-Ne}}$
and $C_V^* = C_V/k_B$.

Standard Monte Carlo simulations of clusters based on the sampling algorithm
proposed by Metropolis {\em et al.}\cite{MRT2} are known to suffer from
systematic errors due to quasiergodicity,\cite{Val-Whit} the non-ergodic
sampling that typically arises with configuration spaces that are comprised
of several regions separated by large barriers. For certain temperature
domains, this can lead to bottlenecks that effectively confine the sampling
to only some of the regions, resulting in large errors.\cite{Allen-Tild}
J-walking addresses the problem of quasiergodicity by coupling the
usual small-scale Metropolis moves to occasional large-scale jumps
that move the random walker to the different regions of configuration space
in a representative manner.\cite{J-walker} The large-scale jumps are governed
by Boltzmann distributions generated at higher temperatures where the
sampling is ergodic. As in my previous study of Ar-Kr clusters, the
more common external file implementation of J-walking was used, with the
J-walker distributions generated beforehand and  stored in an external array
for subsequent jump attempts.\cite{computers}

I also found it necessary to again run several shorter, independent J-walker
trials, instead of the single longer run that is usually done for homogeneous
clusters. The heat capacity results were very sensitive to small errors in
the J-walker distributions, especially at the lower temperatures. These
errors were a consequence of the large increase in the number of isomers
arising from the extra permutational isomers in the binary cluster case,
which made obtaining representative samples that much more difficult.
Therefore, I ran at least five separate J-walking trials, with each trial
sampling from its own unique set of J-walker distributions, to ensure there
were no significant systematic errors in the binary cluster heat capacity
curves. For those clusters showing especially large deviations in their heat
capacity curves at low temperatures, I ran ten trials to reduce the
uncertainties in the peak locations and heights. The results for the separate
trials were then combined and averaged, with the standard deviations taken as
estimates of the uncertainty. As was done with the Ar-Kr calculations, the
total walk length for each temperature was set at $10^6$ passes of data
accumulation. Fig.~\ref{Fig:multi_trials} shows the results for
Ne$_{11}$Ar$_2$. As can be seen in the plot, the noise level in each curve is
about the same magnitude as the differences between the curves over most of
the temperature range, indicating that these walk lengths were sufficiently
long for the desired level of accuracy. However, for the smaller heat
capacity peak occurring at 0.4~K, the curves show a much greater variation,
comparable to the peak height itself. The results for this cluster were the
worst of the series, and while none of the other Ne-Ar clusters showed such
large deviations, the deviations were mostly larger than those obtained for
their Ar-Kr counterparts.

I again ran standard Metropolis simulations for each cluster. The Metropolis
results provided a check of the J-walking results for those temperature
regions where quasiergodicity was not a problem, and they provided additional
insights into the underlying behavior for those regions where quasiergodicity
was evident. Simulations for each temperature consisted of $10^5$ warmup
passes, followed by $10^7$ passes with data accumulation. The temperature
scans were started at the lowest temperature from the global minimum
configuration, with the final configuration for each temperature then used as
the initial configuration for the next temperature. The scans were
continued well past the cluster melting region. Figure~\ref{Fig:multi_trials}
also shows the Metropolis results for Ne$_{11}$Ar$_2$. As can be seen,
substantial discrepancies due to quasiergodicity are evident in the low
temperature region. The Metropolis curve was too low until about 6~K, after
which it rose sharply to approach the J-walker curves. As a check, I repeated
some of the points using walk lengths of $10^8$ passes of data accumulation
per point. Even with these long walks, the Metropolis method was unable to
give accurate results below 6~K, although the results for $T > 6$~K were in
good agreement with the J-walker results (the results obtained for $10^7$
passes remained slightly too low until about 9~K).

The inadequacy of the standard Metropolis algorithm for simulating
heterogeneous clusters at very low temperatures has been noted previously.
Tsai, Abraham, and Pound\cite{TAP} developed a simple but effective strategy
to help overcome these limitations by incorporating an atom exchange scheme
in their Metropolis algorithm. The method enhances the mixing of clusters by
occasionally attempting an exchange move where one of the atoms of one
component is swapped with one of the atoms of another component. The
atom-exchange method showed considerable improvement over the standard
Metropolis method for Ar-Kr clusters in the very low temperature region.
It was even able to partially reproduce the mixing-anomaly peaks that
were completely absent in the Metropolis results, although the agreement with
the J-walking results was only qualitative in these regions. I also ran
atom-exchange simulations for all the Ne-Ar clusters. These were done in a
manner similar to the standard Metropolis runs described previously, with an
atom-exchange move between randomly selected Ne and Ar atoms attempted once
after every pass of standard Metropolis moves. The results for
Ne$_{11}$Ar$_2$ can also be seen in Fig.~\ref{Fig:multi_trials}. In this
case, the atom-exchange method did not provide any improvement over the
standard Metropolis method.

At the lower temperatures associated with the cluster solid-liquid transition
regions, the competition between the intermolecular forces and atomic sizes
effects which isomeric structures play dominant roles in the cluster's
behavior. Thus, information concerning the relative populations of the lower
energy isomers encountered during a simulation at a given temperature can
offer much insight into the nature of cluster dynamics.\cite{SW,Amar_Berry}
The relative frequencies of occurrence for each isomer can be obtained from
quench studies done during the simulation, where the assumption is that the
distribution of minima found in the quenches indicates the likelihood of the
system being associated with a particular isomer. That is, the quench
distributions are a reflection of the relative phase volume of each catchment
basin.\cite{WB} To obtain further insight into the role cluster structure
plays, I also performed quench studies on each of the clusters simulated.
Steepest-descent quenches\cite{SW} were performed periodically on each
cluster during one of its five J-walker trials to monitor the relative
frequency of occurrence for the lower energy isomers as functions of
temperature. For each temperature, quenches were undertaken every 1000
passes, providing 1000 quenched configurations for subsequent analysis. In
each instance, quench trajectories were run until the relative energy
difference converged to within $10^{-7}$.

Having a complete listing of all the stable low-energy isomers for a given
cluster would also be valuable. Although the local minima comprising a
typical cluster potential hypersurface are far too numerous to be completely
catalogued in any practical manner, a reasonably complete listing of the
important lower energy isomers can be obtained easily from the J-walker
distributions. Since the distributions generated at a given temperature
contain representative samples of a cluster's configuration space, a crude,
but nonetheless effective, way to identify cluster isomers is to simply
quench a sufficient number of the configurations stored in the distributions,
saving them in an external file indexed by their energy, and then removing
the duplicate configurations. This was done for all the clusters examined in
this study. Again, each configuration was quenched until the relative energy
difference converged to within $10^{-7}$; the final composite file of unique
isomers was then further refined by running another set trajectories until
the relative energy difference converged to within $10^{-12}$. For 13-atom
Ar-Kr clusters, the low energy isomers in the quenched distributions were
dominated by icosahedral-like structures having the smaller Ar atom in the
center. This was not the case for the Ne-Ar clusters, though. Ne$_{12}$Ar and
Ne$_{11}$Ar$_2$ both had Ne-core icosahedral-like structures as their
lowest-energy isomers, but for most of the other clusters, almost all of the
expected Ne-core icosahedral-like permutational isomers were absent from the
quenched distributions. This implies that these isomers played only minor
roles in the Ne-Ar cluster behavior, which was surprising considering the
dominant role played by the Ar-core icosahedral-like isomers in the Ar-Kr
clusters' behavior. So, for sake of comparison, I determined the
icosahedral-like isomers for each Ne-Ar cluster by generating all the
possible permutations from an icosahedral structure, and then minimized each
one using the quench routines, as described above.

% Ne-Ar Cv paper: results section
\section{Results and Discussion}                        \label{Sec:results}
The structural and thermodynamics results obtained for 13-atom Ne-Ar clusters
were very much different than those obtained for their Ar-Kr
counterparts.\cite{ArKr_Cv} The small difference between the Ar and Kr atom
sizes led to the Ar-Kr clusters essentially having topological configurations
that were very similar to their homogeneous counterparts, but with many
additional permutational isomers arising from the different arrangements of
the Ar and Kr atoms within each type of topological form. As a consequence,
the behavior of 13-atom Ar-Kr clusters was mostly similar to that of their
homogeneous counterpart, and each cluster in the series still retained the
characteristic magic number behavior. Like the Ar$_{13}$ heat capacity curve,
the Ar-Kr heat capacity curves were dominated by a huge peak in the
solid-liquid transition region. Variations within the series arose primarily
because of the energy differences between the permutational icosahedral-like
isomers having an Ar atom core and the icosahedral-like isomers having a Kr
core. For each cluster, the lowest energy isomer corresponded to a segregated
Ar-core icosahedral-like isomer, with the other more mixed Ar-core
icosahedral-like permutational isomers having only slightly higher energies.
These closely spaced isomers gave rise to a very small, low-temperature peak
in the heat capacity for those clusters having more than one such isomer. The
Kr-core icosahedral-like permutational isomers' energies likewise formed a
narrow group. The energy gap between the two sets of permutational isomers
was very large for Ar$_{12}$Kr, but decreased markedly across the series as
the number of Kr atoms increased. This resulted in initially small changes in
the low-temperature side of the solid-liquid transition region heat capacity
peak that grew to eventually form an additional, smaller heat capacity peak
for ArKr$_{12}$. For 13-atom Ne-Ar clusters though, the large size difference
between the Ne and Ar atoms ($\sigma_{\mbox{\scriptsize Ar-Ar}} /
\sigma_{\mbox{\scriptsize Ne-Ne}} = 1.239$ compared to
$\sigma_{\mbox{\scriptsize Kr-Kr}} / \sigma_{\mbox{\scriptsize Ar-Ar}} =
1.062$) resulted in many new non-icosahedral topological forms having no
homogeneous counterparts. These dominated the solid-liquid transition region
for many of the clusters, especially those consisting mostly of Ne atoms,
drastically effecting their resultant heat capacity behavior.

\subsection{Structural properties}
It is those few isomers having the lowest energies that often dominate the
solid-liquid transition region in small clusters. For example, quench studies
of Ar$_{13}$ have shown that at the solid-liquid transition region heat
capacity peak temperature, the lowest 4 isomers accounted for about 80\%
of the quenched configurations.\cite{ArKr_Cv} Thus much insight can be
obtained by examining a cluster's low-lying isomers and their energy
distribution. Figure~\ref{Fig:Minima} shows the 13 lowest energy isomers
found for each of the Ne-Ar clusters; their energies are listed in
Table~\ref{Tbl:Minima}. Unlike the low-lying isomers for Ar-Kr, which were
all readily recognizable as variations of the 4 most stable 13-atom
homogeneous cluster configurations, the low-lying Ne-Ar isomers are mostly
non-icosahedral. The isomers depicted were all obtained from quenches of
J-walker distribution configurations, except for the Ne-core icosahedral-like
isomers for Ne$_9$Ar$_4$ (fifth lowest), Ne$_7$Ar$_6$ (second lowest),
Ne$_4$Ar$_9$ (eighth lowest), and NeAr$_{12}$ (second lowest), which were not
found in the quenched distributions and so were obtained instead by quenching
the permutations derived from icosahedral structures. In fact, very few of
the Ne-core icosahedral-like permutational isomers predicted by the
icosahedral atom-counting rules\cite{TZKDS} were found in the quenched
J-walker distributions, unlike the case for 13-atom Ar-Kr clusters, where all
of the expected permutational icosahedral-like isomers were found.

The lowest energy isomer found for Ne$_{12}$Ar was icosahedral-like,
consisting of a 12-atom Ne sub-cluster stacked on the Ar atom; the
remaining isomers consisted of the small Ne atoms wrapped around the larger
Ar core in various monolayer forms. Although the large difference between the
Ne-Ar and the Ne-Ne potentials favors monolayer structures, the most
efficient Ne packing is in the stacked structure, and in this case, the
packing efficiency overshadowed the differential strengths of the potentials
by a fairly large margin, with the difference between the stacked
configuration and the lowest monolayer configuration being $1.31042
\epsilon_{\mbox{\scriptsize Ne-Ne}}$. The stacked and monolayer forms are
reminiscent of the wetting and nonwetting structures observed by Chartrand,
Shelley and Le~Roy\cite{CSLR} in their study of \mbox{SF$_6$--(rare~gas)$_n$}
clusters. They observed both stacked (or nonwetted) and monolayer (wetted)
structures for all the clusters $n \geq 4$ for SF$_6$-Ar$_n$ and $n \geq 5$
for SF$_6$-Kr$_n$, with wetted isomers having the lowest energy for some $n$,
and nonwetted isomers for the others. For $n = 12$, the ground state
configuration for both Ar and Kr had the larger SF$_6$ located on the
exterior of a stacked 12-atom sub-cluster, while in the second-lowest energy
configuration, the SF$_6$ core was surrounded by a monolayer of rare gas
atoms. For SF$_6$-Ar$_{12}$, the energy difference was
$1.544\epsilon_{\mbox{\scriptsize Ar-Ar}}$, which is similar to the
Ne$_{12}$Ar energy difference in the analogous reduced units. Note also, that
the size difference between the Ne and Ar atoms is similar to the size
difference between Ar and SF$_6$. For the SF$_6$-Rg$_n$ clusters, pentagonal
and fused pentagonal caps were especially stable configurations that appeared
often in the isomers for $n \geq 6$, and these are likewise evident in many
of the Ne$_{12}$Ar isomers shown in Fig.~\ref{Fig:Minima}.

For Ne$_{11}$Ar$_2$, there are 3 Ne-core icosahedral-like structures
possible. The lowest energy isomer found consisted of an icosahedral-like
structure with the 2 Ne atoms occupying adjacent sites, analogous to the
Ar$_{11}$Kr$_2$ lowest energy isomer. But unlike Ar$_{11}$Kr$_2$, there were
also two non-icosahedral isomers having energies lower than the other two
icosahedral-like permutational isomers. These non-icosahedral
isomers are characterized by a 6-atom ring wrapped around the axis formed by
the 2 Ar atoms. The other 5 Ne atoms occupied the various interstices formed
by this Ne$_6$Ar$_2$ sub-cluster. The lowest energy non-icosahedral
configuration had the 5 Ne atoms forming a partial second ring just beneath
the 6-atom ring. This arrangement can also be seen to form two fused
pentagonal caps, each consisting of one of the Ar atoms and 4 Ne atoms capped
by a fifth Ne atom. This is a relatively stable configuration, so unlike the
Ne$_{12}$Ar case, the energy difference between the lowest energy stacked and
monolayer configurations was very small ($0.0537\epsilon_{\mbox{\scriptsize
Ne-Ne}}$).

The increase in the Ne-core icosahedral-like isomer energies relative to the
non-icosahedral isomers continued for Ne$_{10}$Ar$_3$, which has 5 such
isomers. The lowest energy icosahedral-like isomer (analogous to the ground
state Ar$_{10}$Kr$_3$ isomer) was not the ground state isomer, but ranked
third lowest; the second-lowest Ne-core isomer ranked eleventh. Unlike the
Ar-Kr clusters, the 5 Ne-core icosahedral-like isomers were not closely
spaced in energy. Their energy range was $1.74461 \epsilon_{\mbox{\scriptsize
Ne-Ne}}$, compared to $0.12115 \epsilon_{\mbox{\scriptsize Ar-Ar}}$ for the 5
analogous Ar$_{10}$Kr$_3$ Ar-core isomers. The non-icosahedral configurations
shown in Fig.~\ref{Fig:Minima} each consist of the 3 Ar atoms arranged in a
trigonal plane having 3 of the Ne atoms stacked on them in an opposing
trigonal plane, and the remaining Ne atoms occupying various sites around
them. The lowest energy configuration has C$_{3v}$ symmetry.

The lowest energy isomer for Ne$_9$Ar$_4$ also consisted of the Ar atoms
arranged in the lowest energy homogeneous sub-cluster configuration (the
tetrahedron in this case), with the Ne atoms occupying the exterior sites.
Its structure was very similar to the second lowest energy Ne$_{10}$Ar$_3$
isomer, with the main difference being the replacement of the Ne atom
directly behind the 3 Ar atoms by another Ar atom. Similarly, the second
lowest energy Ne$_9$Ar$_4$ isomer had a structure akin to the ground state
structure of Ne$_{10}$Ar$_3$, with the main difference again being the
replacement of the Ne atom directly behind the 3 Ar atoms by another Ar atom,
so that the structure also has C$_{3v}$ symmetry. The other Ne$_9$Ar$_4$
isomers can also be obtained formally from similar Ne$_{10}$Ar$_3$ structures
by replacing a Ne atom with an Ar atom. Ne$_9$Ar$_4$ also has 10 Ne-core
icosahedral-like isomers. The lowest energy one (analogous to the ground
state Ar$_9$Kr$_4$ isomer) ranked fifth lowest. The second and fourth lowest
energy Ne$_9$Ar$_4$ icosahedral-like isomers were found in the quench
distribution, but none of the other 7 were. The reason for the absence of the
Ne$_9$Ar$_4$ icosahedral-like isomers in the quenched distributions was
evident in the quench trajectories obtained from one of the J-walking walks:
almost all of the very few lowest energy Ne-core icosahedral-like isomers
encountered during the walk quenched down to the global minimum, implying
that the barriers to rearrangement between the two isomers are low. Thus,
while the quenched J-walker distributions did not contain all of the
low-lying isomers, the manner in which they were obtained suggests that they
did contain the most important stable isomers, and that the shortfall
consisted primarily of those isomers having low likelihood and being easily
rearranged. Again, the energy spread between the Ne$_9$Ar$_4$ Ne-core
icosahedral-like isomers was very large ($2.34926\epsilon_{\mbox{\scriptsize
Ne-Ne}}$ compared to $0.16439\epsilon_{\mbox{\scriptsize Ar-Ar}}$ for
Ar$_9$Kr$_4$).

The lowest energy configuration for a 5-atom Lennard-Jones cluster is the
trigonal bipyramid,\cite{Hoare-Pal} and so it not surprising that the
lowest energy isomer for Ne$_8$Ar$_5$ consisted of the 5 Ar atoms arranged in
a trigonal bipyramid configuration, with the 8 Ne atoms attached to the
exterior sites, resulting in C$_{2v}$ symmetry. The other low-lying
non-icosahedral isomers had a similar form, with some of the Ne atoms
occupying other exterior locations. The lowest energy configuration for a
6-atom Lennard-Jones cluster is the octahedron, but the tripyramid is only a
little higher in energy ($-12.30293\epsilon$ compared to $-12.71206\epsilon$)
and the two are connected by low saddle points.\cite{Hoare-Pal} As can be
seen in Fig.~\ref{Fig:Minima}, the lowest energy non-icosahedral isomers
found for Ne$_7$Ar$_6$ all have their 6 Ar atoms arranged in the tripyramid
form, and not the lower energy octahedron. The lowest energy octahedral
isomer, which has each Ne atom located on one of the octahedral faces, was
the 36th lowest energy isomer found, with an energy of
$1.63820\epsilon_{\mbox{\scriptsize Ne-Ne}}$ above the global minimum.
Although the Ar-Ar interactions are maximized slightly in the octahedral
arrangement, the Ne-Ne interactions are reduced substantially, making such
configurations unfavorable. Ne$_7$Ar$_6$ has 18 Ne-core icosahedral-like
isomers, and the lowest energy one ranked second; it was not found in
the quenched distributions, and only two of the other 17 isomers were found.
According to the icosahedral atom-counting rules,\cite{TZKDS} Ne$_7$Ar$_6$ is
also expected to have 12 Ar-core icosahedral-like isomers, but because the Ne
atoms are too small to remain fixed in the icosahedral sites around the much
larger Ar core atom, many additional isomers resembling highly distorted
icosahedral-like forms are possible. Most of these were found in the quenched
distributions. Their energies were generally higher than the Ne-core
icosahedral-like isomer energies, but because of the large energy spread in
both sets of isomers, there was substantial overlap. The lowest-energy
Ar-core icosahedral-like isomer had an energy of
$1.99495\epsilon_{\mbox{\scriptsize Ne-Ne}}$ relative to the global minimum.

The 7-atom Lennard-Jones cluster is one of the magic number sizes, having a
pentagonal bipyramidal structure for its lowest energy isomer. Most of the
Ne$_6$Ar$_7$ low energy isomers shown in Fig.~\ref{Fig:Minima} consist of
the 6 Ne atoms located on the trigonal faces of an Ar-atom pentagonal
bipyramidal sub-cluster. The lowest energy isomer has 2 pairs of Ne atoms
located on the upper trigonal faces, and the other pair located on the lower
face below. The pentagonal sites formed by the Ar-atom ring are separated by
distances much larger than the Ne-Ne equilibrium distance, and so the
distances between the Ne atoms in each pair are compressed substantially. The
second lowest energy isomer for Ne$_6$Ar$_7$ is readily recognizable as a
truncated Ar-core icosahedral-like structure, with one of the Ne atoms
located on one of the Ar icosahedral faces. If the
Ne atom were moved to the top to cap the 5-atom Ne ring, the resulting
configuration would be the lowest energy Ar-core icosahedral-like isomer.
This isomer ranked 24th lowest, with an energy of
$1.11047\epsilon_{\mbox{\scriptsize Ne-Ne}}$ above the ground state. The
energy gained by increasing the number of Ne-Ar interactions in the truncated
icosahedral-like form is much greater than the energy lost by breaking the
compact icosahedral structure. The third lowest energy isomer did not have
its Ar atoms arranged as a trigonal bipyramid, but as an incomplete stellated
tetrahedron, the second lowest 7-atom Lennard-Jones isomer.\cite{Hoare-Pal}

Ne$_5$Ar$_8$ represents an interesting cross-over point in the 13-atom Ne-Ar
series. It is the last in the series having a non-icosahedral structure as
its lowest energy isomer --- Ne$_4$Ar$_9$ to NeAr$_{12}$ all have Ar-core
icosahedral-like isomers as their ground state isomers. Thus, it is also the
last in the series to have its lowest energy Ne-core icosahedral-like isomer
being more stable than its lowest energy Ar-core icosahedral-like isomer. The
lowest energy isomer consists of the 8 Ar atoms arranged as a fully
stellated tetrahedron, with the 5 Ne atoms located on the tetrahedral faces.
The stellated tetrahedron is the second lowest energy isomer for 8-atom
homogeneous Lennard-Jones clusters (the pentagonal bipyramid with an added
atom is the lowest energy isomer).\cite{Hoare-Pal} The energy difference
between the two is not small ($-19.82149\epsilon$ compared to
$-18.97606\epsilon$), but the larger number of Ne-Ar nearest neighbor
interactions in this conformation is sufficient to overcome the deficit.
Except for the third lowest energy isomer, which is a Ne-core
icosahedral-like structure, all the other Ne$_5$Ar$_8$ isomers depicted in
Fig.~\ref{Fig:Minima} consist of the Ar atoms arranged in the lowest energy
8-atom homogeneous sub-cluster. The second lowest energy isomer is a
truncated Ar-core icosahedral-like structure, having one of the Ne atoms
occupying one of the icosahedral faces formed by 3 Ar atoms. This structure
is only slightly more stable than the lowest energy Ar-core icosahedral-like
isomer, which ranked tenth lowest, but only
$0.12235\epsilon_{\mbox{\scriptsize Ne-Ne}}$ higher in energy.

For Ne$_4$Ar$_9$, the two lowest energy isomers were Ar-core icosahedral-like
permutational isomers, and so this cluster was the first in the series to
have Ar-core icosahedral-like isomers that were lower in energy than their
corresponding truncated icosahedral-like isomers. The energy ranges of the
icosahedral-like and truncated icosahedral-like permutational isomers were
both larger than the energy difference between the two sets, and so the
isomers were mostly interspersed. As the number of Ar atoms increased, so
too did the energy difference between the two sets, and they became
increasingly segregated. For Ne$_3$Ar$_{10}$, the 4 lowest energy isomers
were Ar-core icosahedral-like permutational isomers, and the fifth one ranked
twelfth lowest overall. For Ne$_2$Ar$_{11}$, all 3 Ar-core icosahedral-like
permutational isomers lay well below the truncated icosahedral-like
permutational isomers, and the lone Ne-core icosahedral-like isomer was
fourth lowest. For NeAr$_{12}$, the Ar-core icosahedral-like isomer lay well
below its Ne-core counterpart. This energy was similar to the energies of the
next three truncated icosahedral-like isomers, all of which have the Ne atom
occupying one of the exterior sites. The subsequent isomers were also
separated from these isomers by nearly as large an energy gap.

The trends in the lowest energy icosahedral-like and non-icosahedral isomers
as functions of the cluster composition can be seen in
Fig.~\ref{Fig:MinimaComp}. The reduced energies show a nearly linear
dependence on composition that is due to the large difference between the Ne
and Ar intermolecular forces. As the fraction of Ar atoms increased, so
too did the fraction of stronger Ar-Ne and Ar-Ar interactions. This effect
overshadowed the smaller, more subtle trends, which can be made manifest
by scaling out the gross linear dependence on composition by using energies
scaled by the composition weighted average $\epsilon_{\rm Avg} = X_{\rm Ne}
\epsilon_{\mbox{\scriptsize Ne-Ne}} + X_{\rm Ar} \epsilon_{\mbox{\scriptsize
Ar-Ar}}$, where $X_{\rm Ne} = n_{\rm Ne}/(n_{\rm Ne} + n_{\rm Ar})$ and
$X_{\rm Ar} = 1 - X_{\rm Ne}$. The icosahedral-like isomers' scaled energies
increased as their mole fraction decreased from unity. For the Ne-core
isomers, they rose rapidly from Ne$_{13}$ until they leveled off at
Ne$_9$Ar$_4$, and then decreased slightly again after Ne$_6$Ar$_7$. The
Ar-core scaled energies likewise increased from the Ar$_{13}$ value as the
number of Ne atoms increased. However, they did not level off, but kept
rising in a roughly linear manner. The two Ne-Ar icosahedral-like curves
crossed at Ne$_5$Ar$_8$. These trends are strongly dependent on the atomic
sizes, though. Had the Ne and Ar atoms been the same size, then the Ar-core
isomers would have been more stable than the Ne-core isomers for all
compositions, since the central Ar atom would have maximized the Ne-Ar
interactions relative to the weaker Ne-Ne interactions. For 13-atom Ar-Kr
clusters, the size difference between the Ar and Kr atoms was sufficiently
large that the trends reversed and the Kr-core isomers were higher in energy
than the Ar-core isomers for all compositions. The Ne-Ar curves demonstrate
that there exists another critical size ratio for those icosahedral-like
isomers consisting mostly of the larger-atom component, where having the
larger atom at the center becomes more stable again than having the smaller
atom at the center. The scaled energies for the non-icosahedral isomers were
roughly constant across the series. The similarity of the scaled energies for
these isomers indicates that these clusters had non-icosahedral ground state
structures not because the isomers themselves became intrinsically more
stable as the composition became more mixed, but because the icosahedral-like
isomers became so much less stable.

The effects of cluster composition on the isomer energy distributions are
shown in Fig.~\ref{Fig:Min_Energies}, which depicts the energy spectrum for
each cluster's local minima relative to its global minimum. The plots have
been arranged so that complementary clusters (Ne$_{13-n}$Ar$_n$ and
Ne$_n$Ar$_{13-n}$) appear in the same column. As was the case with 13-atom
Ar-Kr clusters, the densities of the energetically distinct local minima
increased as the number of possible permutations increased, quickly becoming
so great that most of the individual levels cannot be distinguished in the
plots. Otherwise, though, the plots differ greatly from the Ar-Kr plots. In
those plots, the clusters having a greater fraction of Ar atoms had an energy
spectrum similar to that of Ar$_{13}$, which is dominated by its very large
energy gap between the icosahedral ground state and the non-icosahedral
isomers. As the fraction of Kr atoms in the clusters increased from
Ar$_{12}$Kr to Ar$_7$Kr$_6$, the number of Ar-core icosahedral-like
permutational isomers increased, forming a small group of closely spaced
low-lying energies, but the energy gap between these isomers and the band
formed by the higher energy truncated icosahedral and non-icosahedral isomers
remained nearly constant. The Kr-core icosahedral-like isomers were also
closely spaced, but the gap between them and the Ar-core isomers decreased
steadily as the Kr mole fraction increased. The Ne-Ar energy spectra do not
exhibit such patterns, showing instead, a wide variety of behavior. Except
for NeAr$_{12}$, none of the binary cluster spectra has the large gap between
the global minimum and the higher energy isomers that characterize the
homogeneous cluster spectra. The number of low lying isomers ($\Delta E
\lesssim 2$) is greater for those clusters having similar numbers of Ne and
Ar atoms, being greatest for Ne$_5$Ar$_8$. For these clusters, the low-lying
icosahedral-like and non-icosahedral isomers have similar energies so that
the two sets are substantially interlaced.

\subsection{Thermodynamic properties}
The wide variety that characterizes the structural properties of 13-atom
Ne-Ar clusters carries over to their thermodynamic properties as well.
Figure~\ref{Fig:Cv} shows the heat capacity (in reduced units) for each
cluster as a function of temperature. The heat capacity peak heights and
temperatures are listed in Table~\ref{Tbl:Cv_peaks}, and the values for the
major peaks are plotted in Fig.~\ref{Fig:Cv_Peaks}. The J-walking curves
shown in each plot in Fig.~\ref{Fig:Cv} represent the averages obtained from
combining the results of the individual J-walking runs, as described in
Section~\ref{Sec:theory}. Representative standard deviations have not been
included to prevent excessive cluttering, but they were consistent with the
noise levels evident in the curves. The peak values were obtained in each
case from the averaged J-walker curves. These curves were smoothed and then
interpolated in the peak vicinities to obtain finer mesh sizes, and the peak
parameters were then found simply by searching the interpolated
data.\cite{Sav_Gol} The uncertainties in the peak heights were estimated from
the average standard deviations of the points in the vicinity of each peak.

The plots of the Ne-Ar heat capacity peak temperature and height as functions
of composition shown in Fig.~\ref{Fig:Cv_Peaks} do not exhibit the smooth
variations that were seen in the corresponding plots of their Ar-Kr
counterparts. The Ne-Ar peak temperatures can be grouped into three regions:
Ne$_{13}$ to Ne$_{10}$Ar$_3$, which have similar, but much lower peak
temperatures than the others; Ne$_9$Ar$_4$ to Ne$_4$Ar$_9$, whose peak
temperatures are roughly constant; and Ne$_3$Ar$_{10}$ to Ar$_{13}$, whose
peak temperatures rise nearly linearly as a function of the composition. The
linear rise in the peak temperature is similar to the trends observed for the
Ar-Kr clusters, which showed a gradual linear rise from Ar$_{13}$ to
Ar$_7$Kr$_6$, followed by a steeper linear rise from Ar$_6$Kr$_7$ to
Kr$_{13}$. The changes in the Ar-Kr peak heights were relatively minor, with
the difference between the maximum and minimum being only about 12\% of the
Ar$_{13}$ peak height. The changes in the Ne-Ar peak heights were much more
drastic. Again, it was those Ne-Ar clusters having a large Ar mole fraction
whose behavior most resembled that of their Ar-Kr counterparts. Also shown in
Fig.~\ref{Fig:Cv_Peaks} is a plot of the peak heights as a function of the
peak temperature. Unlike the corresponding plot for the Ar-Kr clusters,
though, which consisted of a regular progression on a smoothly varying curve,
the Ne-Ar plot consists of an irregular scatter of points with some grouping
evident.

Except for the NeAr$_{12}$ heat capacity curve, none of the curves shown in
Fig.~\ref{Fig:Cv} resembles the Ar$_{13}$ curve obtained
previously,\cite{Magic_Cv,J-walker} unlike the Ar-Kr curves, which were
mostly very similar to the Ar$_{13}$ curve in the solid-liquid transition
region. The similarity in the Ar-Kr curves was a consequence of the Ar-core
icosahedral-like permutational isomer energies being closely spaced and well
separated from the Kr-core icosahedral-like energies, and from the
non-icosahedral energies, so that the important differences occurred only in
the low temperature regions. Except for NeAr$_{12}$, the Ne-core
icosahedral-like, Ar-core icosahedral-like, and the non-icosahedral isomer
energies were not well separated, but were mostly intermixed, so that there
were few large gaps in the energy spectra. This resulted in the much
diminished heat capacity peaks evident in Fig.~\ref{Fig:Cv}.

The standard Metropolis heat capacity curves for each cluster have also
been included in Fig.~\ref{Fig:Cv}. These were in good agreement with the
J-walking curves over most of the temperature domain for only a few of the
clusters, but were in mostly good agreement with all of them at the higher
temperatures. The poor results obtained from the standard Metropolis method
in the low temperature regions was not surprising since this was evident in
the 13-atom Ar-Kr Metropolis results as well, where the low temperature
mixing-anomaly peaks were not reproduced at all. What was surprising is how
little improvement was obtained using the atom-exchange method. For the Ar-Kr
clusters, the atom-exchange results were substantially better than the
standard Metropolis results, and were in good agreement with the J-walker
results for all the clusters over most of their temperature domains. The
atom-exchange results even showed the mixing-anomaly peaks, although they
were only in qualitative agreement with the J-walker results. As is evident
in Fig.~\ref{Fig:Cv}, the atom-exchange results were mostly similar to the
standard Metropolis results for most of the Ne-Ar clusters. The poor
performance of the atom-exchange method in application to Ne-Ar clusters is
also due to the large size difference between the Ne and Ar atoms. In the low
temperature regions where most of the Metropolis configurations have energies
only slightly above their nearest local minimum, the trial configurations
most often obtained by swapping one of the large Ar atoms with a small Ne
atom were strained configurations having energies too high to be accepted
with sufficient frequency. Comparison of the exchange-move acceptance ratios
for the Ne-Ar clusters with the ratios for their corresponding Ar-Kr
counterparts showed that indeed far fewer exchange moves had been accepted
for the Ne-Ar clusters. It appears then, that the atom-exchange method is
useful for reducing quasiergodicity in binary cluster simulations only when
the atomic sizes of the two components are similar enough.

Much of the low temperature Ne-Ar heat capacity behavior seen in
Fig.~\ref{Fig:Cv} can still be rationalized mainly in terms of the potential
minima energy spectra shown in Fig.~\ref{Fig:Min_Energies}, although there
are other important factors involved as well. The curves for the
predominately Ne clusters Ne$_{12}$Ar, Ne$_{11}$Ar$_2$ and Ne$_{10}$Ar$_3$
differed from the others by their much lower peak temperatures, and by their
much smaller peak sizes. The peak temperatures were all substantially below
the Ne$_{13}$ peak temperature of 10.15~K. Ne$_{12}$Ar had a single
moderately high peak occurring at 5.20~K that was evident only in the
J-walking calculations; the Metropolis and atom-exchange results showed an
abrupt rise in the heat capacity at 6~K, but no peak. The peak is due to the
moderately large energy gap between the stacked ground state icosahedral-like
isomer and the Ar-core monolayer isomers. The rest of the curve rising
sharply after the peak is the low temperature side of a larger peak
associated with the cluster's dissociation.\cite{DissPeak} It is tempting to
extend the correspondence between the Ne$_{13}$ heat capacity peak and its
solid-liquid transition region to Ne$_{12}$Ar and to think of its peak then
as characterizing a solid-liquid transition between the ``solid'' stacked
configuration and the ``liquid'' monolayer configurations, but this is not
the case. The absence of a peak in the Metropolis results implies that this
peak is not due to any dynamic transition between the stacked and monolayer
isomers, and so cannot represent a coexistence region like the one in
Ne$_{13}$. Instead, once the Metropolis simulations reached 6~K, the
Ne$_{12}$Ar clusters began achieving energies high enough to overcome the
potential barriers separating the stacked configurations from the other
low-lying isomers, whereupon they escaped their confinement, transformed into
one of the monolayer configurations and were seldom encountered again. As
will be discussed later, the quench studies revealed that as the temperature
continued increasing, it was the monolayer isomers that dominated the
solid-liquid transition region, which occurred at higher temperatures.

For Ne$_{11}$Ar$_2$, the Ne-core icosahedral-like ground state and the lowest
energy non-icosahedral isomer have very similar energies, which resulted in
the very small heat capacity peak occurring at 0.37~K. As with Ne$_{12}$Ar,
there was a very large discrepancy between the J-walking and Metropolis
results over much of the low temperature region. The Metropolis curve rose
slowly and nearly linearly until about 6~K, after which it rose abruptly to
approach the J-walking curve. The curves form a larger peak at 8.92~K that
can be identified, in conjunction with the quench studies discussed later, as
the solid-liquid transition region. Ne$_{10}$Ar$_3$ had a peak that was more
pronounced than the peak for Ne$_{11}$Ar$_2$, and it was located at a
slightly lower temperature of 6.76~K. The Metropolis and atom-exchange curves
for Ne$_{10}$Ar$_3$ were in much better agreement with the J-walking curve
than was the case for the other two clusters. Although the Metropolis peak
was much diminished and the curve on the low temperature side of the peak was
much lower than the J-walking curve, the results were at least in qualitative
agreement.

The predominately Ar clusters Ne$_3$Ar$_{10}$ to NeAr$_{12}$ were
distinguished by a single large peak corresponding to the solid-liquid
transition region. The peaks became smaller and shifted to lower
temperatures as the number of Ar atoms decreased. Their larger size is due to
the very large gaps between the lowest energy isomers and the band of higher
energy isomers that are accessible only at the higher temperatures associated
with the liquid region. As the gap decreased from Ar$_{13}$ to
Ne$_3$Ar$_{10}$, so too did the peak height. These clusters also have
sufficiently large energy gaps between the ground state isomer and the next
lowest energy isomer that none exhibited the small, very low temperature heat
capacity peaks seen in some of the other clusters. While the heat capacity
curve for NeAr$_{12}$ most resembled the Ar$_{13}$ curve, it differed
much from its ArKr$_{12}$ counterpart. The ArKr$_{12}$ curve had a smaller,
well separated second peak occurring just before the solid-liquid transition
peak that was due to the interaction between the Ar-core and Kr-core
icosahedral-like isomers. The isomer energy spectra for NeAr$_{12}$ and
ArKr$_{12}$ are mostly similar. For ArKr$_{12}$, the Kr-core isomer having
the larger Kr atom at the center was $3.15883\epsilon_{\mbox{\scriptsize
Ne-Ne}}$ higher in energy than the Ar-core isomer. For NeAr$_{12}$, the
energy gap between the Ne-core and Ar-core icosahedral-like isomers was only
slightly larger ($3.99438\epsilon_{\mbox{\scriptsize Ne-Ne}}$), but in this
case, it was the Ar-core isomer with its larger Ar atom at the center that
had the lower energy. NeAr$_{12}$ also differs from ArKr$_{12}$ in that its 3
truncated icosahedral-like isomers with the Ne atom located on one of the
icosahedral faces had energies only slightly higher than the Ne-core
icosahedral-like isomer; in ArKr$_{12}$, the corresponding truncated
icosahedral-like isomers had much higher energies, with the gap being similar
to the one in Kr$_{13}$. Otherwise though, the two energy spectra differed
little. This major difference between the NeAr$_{12}$ and ArKr$_{12}$ heat
capacity curves despite their similar energy spectra clearly indicates that
much more than just the isomer energy distributions are involved in the
clusters' thermodynamic behavior. Both NeAr$_{12}$ and Ne$_2$Ar$_{11}$ showed
mostly good agreement between the J-walking and Metropolis curves, while
Ne$_3$Ar$_{10}$ showed significant deviations on the low temperature side of
the peak. For Ne$_3$Ar$_{10}$, the atom-exchange results were markedly better
than the Metropolis results, though.

The heat capacities for the intermediate clusters Ne$_9$Ar$_4$ to Ne$_4$Ar$_9$
are characterized by very broad, low peaks extending over much of the
temperature domain. While the curve profiles transformed smoothly in going
from Ne$_4$Ar$_9$ to Ne$_3$Ar$_{10}$, there was an abrupt change in going
from Ne$_{10}$Ar$_3$ to Ne$_9$Ar$_4$. The peaks for Ne$_9$Ar$_4$ to
Ne$_4$Ar$_9$ occurred at much higher temperatures than did the peaks for
Ne$_{12}$Ar to Ne$_{10}$Ar$_3$, but their temperatures were similar to the
solid-liquid transition peak temperatures obtained for Ne$_3$Ar$_{10}$ to
NeAr$_{12}$. While the similar peak temperatures might suggest that these
peaks also corresponded to solid-liquid transitions, the peaks' broad
profiles, extending over such large temperature ranges, together with the
abrupt changes in the peak height and location between Ne$_{10}$Ar$_3$ and
Ne$_9$Ar$_4$, indicates a more complicated scenario, one which will await the
discussion of the quench results. Except for Ne$_4$Ar$_9$, the peak
temperatures for this group were remarkably similar. The peak heights mostly
decreased over this range as the Ar mole fraction increased, reaching a
minimum at Ne$_5$Ar$_8$. Those clusters having isomers with energies slightly
above their global minimum energy (Ne$_9$Ar$_4$, Ne$_8$Ar$_5$, Ne$_6$Ar$_7$,
and Ne$_5$Ar$_8$) also had an additional small peak occurring at very low
temperatures. In the Ar-Kr clusters, these smaller peaks' sizes were
primarily determined by the number of Ar-core icosahedral-like isomers, since
these isomers were grouped closely together in energy. For the Ne-Ar clusters
though, the large spread of the various permutational isomer energy ranges
implies that there is no such simple connection between the low-lying isomers
and the size of the low temperature peaks: Ne$_9$Ar$_4$ had the largest such
peak, which is due to the transactions between its lowest 3 non-icosahedral
permutational isomers, while Ne$_8$Ar$_5$, with 5 low-lying non-icosahedral
isomers, had a much smaller peak.

The potential energy curves for the clusters as functions of temperature are
shown in Fig.~\ref{Fig:Caloric}. The large difference between the Ne and Ar
intermolecular potentials results in the binary cluster curves having widely
varying energies and temperature ranges, which makes it difficult to compare
them directly. It is clear, though, that they differ qualitatively from the
homogeneous cluster curves; except for the Ne$_{12}$Ar and NeAr$_{12}$
curves, the curves for the binary clusters show very gradual rises occurring
over more extended temperature ranges. The Ne$_{12}$Ar curve has a noticeable
inflection, but it occurs at a reduced temperature well below that of the
Ne$_{13}$ inflection.

\subsection{Quench results}
Further insights into the temperature dependence of binary cluster
isomerizations can be obtained from the quench studies.
Figures~\ref{Fig:Quench1} to \ref{Fig:Quench3} show the quench results
obtained using J-walking for the six lowest-energy isomers for each Ne-Ar
cluster. Not surprising, the quench curves also show much variation, and are
generally dissimilar from the corresponding homogenous cluster quench curves.
This again is very much different than what occurred for the 13-atom Ar-Kr
clusters, whose quench curves showed large differences due to the low-lying
Ar-core icosahedral-like permutational isomers; if those curves were combined
into single curves representative of their common icosahedral-like
topological forms, then the resulting quench profiles were in fact very
similar to their homogeneous counterparts.

For Ne$_{12}$Ar, the fraction of quenches to the ground state Ne-core
icosahedral-like isomer dropped off rapidly after 4~K as the Ar-core
monolayer isomers began to contribute, so that by 8~K, there were very few
quenches to this isomer. The two lowest-energy monolayer isomers dominated
the quenches over much of the rest of the temperature domain depicted,
contributing about 75\% of the quenches at 7~K (which the Metropolis heat
capacity results indicate as being in the solid-liquid transition region),
and about 40\% of the quenches at 13~K, where the large fraction of quenches
to the other higher energy isomers is indicative of liquid-like behavior.
The curve for the remaining higher energy isomers rose much more slowly than
was the case for the other predominately Ne clusters, implying that the
solid-liquid transition region for Ne$_{12}$Ar seems to extend over a much
larger temperature range. Note too, that this curve becomes dominant well
beyond the heat capacity peak, again consistent with the Metropolis heat
capacity behavior indicating that the solid-liquid transition region occurred
at higher temperatures.

To gain more insight into the melting behavior of this cluster, I ran
additional quench studies using similar Metropolis temperature scans in the
direction of decreasing temperatures, starting at 10~K.
Figure~\ref{Fig:QuenchMet} compares the quench profiles obtained for the
three lowest energy isomers from these simulations with those obtained from
the Metropolis simulations in the direction of increasing temperatures, which
were begun from the ground state isomer. The curves show clear hysteresis.
The quench curves in the increasing temperature direction reveal that the
clusters were locked into their lowest energy configuration until 6.3~K,
after which the monolayer isomers began being accessed (this temperature also
corresponds to the point where the Metropolis heat capacity curve rose
abruptly in Fig.~\ref{Fig:Cv}). Reversing the temperature direction at 10~K
resulted in curves that were in agreement only for temperatures higher than
6.3~K, below which, none of the quenches were to the ground state isomer, but
were distributed among the low lying monolayer isomers. This behavior implies
that the barriers to isomerization between the monolayer configurations are
very low, while the barriers between the stacked and the monolayer
configurations are very large.

The spikes seen in the Metropolis quench curves near 7~K imply that this is a
narrow coexistence region where the cluster spent long periods of time in
either the stacked or monolayer configurations, switching between the two
only occasionally. But this coexistence is quite different than the one
occurring in Ne$_{13}$. For Ne$_{13}$, the coexistence region primarily
involved isomerizations between the ground state icosahedron and the three
higher energy truncated icosahedral structures having one of the atoms
located on one of the icosahedral faces (at the heat capacity peak
temperature, about 80\% of the quenches were to these four isomers). It is
the very large energy gap between the Ne$_{13}$ ground state and the
truncated icosahedral isomers, as well as the substantial gap between those
isomers and the succeeding non-icosahedral isomers, that is responsible for
the pronounced coexistence range in Ne$_{13}$. The gap between the
Ne$_{12}$Ar ground state and the corresponding truncated icosahedral-like
isomers having a lone Ne atom on one of the icosahedral faces is slightly
smaller than the one in Ne$_{13}$, but as can be seen in
Fig.~\ref{Fig:Min_Energies}, Ne$_{12}$Ar has a large number of monolayer
isomers with energies well below the truncated icosahedral-like isomers. Thus
once a stacked Ne$_{12}$Ar isomer at about 6.3~K gained sufficient energy to
move a Ne atom from an icosahedral vertex onto an icosahedral face, it
reached a plateau region where the lone Ne atom could return again to its
icosahedral site, or the cluster could continue isomerizing into one of the
monolayer isomers. But once it started down the monolayer path, it could
easily continue isomerizing into several of the various monolayer isomers,
since the barriers separating them are so low. Such a trajectory would cover
a much larger distance in configuration space, greatly reducing the
likelihood of its return to the original stacked configuration. For
Ne$_{13}$, the four lowest energy isomers are close together in configuration
space so that the isomerizations between them are largely reversible, and the
typical recurrence times between isomers returning to icosahedral
configurations are short on the time scale of the walk length. For
Ne$_{12}$Ar, the recurrence times were so much longer that the
transformations from stacked to monolayer configurations were largely
irreversible, except for a very narrow temperature range, where the
probability of crossing the barriers was so low that such an occasional
transformation would leave the cluster trapped in the stacked configuration
subspace for a long period of time before accessing enough energy to take it
out again.

The icosahedral-like configuration subspace being so much smaller
than the non-icosahedral configuration subspace, and it being isolated by
large barriers, were also factors in the quench distributions for the other
Ne-Ar clusters, where the quenches to Ne-core icosahedral-like isomers were
also very low; it also accounts for the absence of many of these
permutational isomers in the J-walker distributions noted earlier. Similar
behavior was encountered by Matro, Freeman and Topper in their study of
ammonium chloride clusters.\cite{Matro_Freeman} The second and third lowest
energy isomers for (NH$_4$Cl)$_3$ were only slightly higher in energy than
the lowest energy isomer ($\Delta E_{01} = 235$ K, $\Delta E_{02} = 635$ K)
and separated from it by large barriers of nearly equal size ($\Delta E_{01}
= 2118$ K, $\Delta E_{02} = 2021$ K). During the simulations at 50~K, all the
configurations were found to be in the potential energy wells belonging to
the first and third isomers, but none were found in the well belonging to the
second isomer, which the authors attributed to the much smaller fraction of
configuration space associated with the second isomer.

Visually examining samples of the Ne$_{12}$Ar configurations generated during
Metropolis walks at various temperatures provided further support for the
above arguments. Walks begun from an initial ground state configuration at
8~K quickly resulted in configurations that were distended, but
quite recognizable as stacked Ne-core icosahedral-like structures. After a
several thousand passes, the cluster isomerized into one of the truncated
icosahedral-like forms, where the lone exterior Ne atom slowly migrated
over the surface, moving from site to site, sometimes returning to its
original site. Eventually, after many additional steps, the exterior Ar atom
and the core Ne atom would make a concerted movement that brought the
Ar atom into the central location and moved the central Ne atom to the
exterior, transforming the cluster into an Ar-core precursor that quickly
collapsed into a monolayer structure, after which the exterior Ne atoms
rapidly rearranged themselves into various other monolayer configurations.
The migration of exterior Ne atoms from one side of the Ar atom to another
was still sluggish, though, because the Ne atoms enveloped the Ar atom
closely enough to hinder gross movements. Repeating the walks
at higher temperatures led to similar behavior, but on an accelerated time
scale such that the transformation from stacked to monolayer configuration
occurred much more quickly. The monolayer configurations also underwent
isomerizations more rapidly as the exterior Ne atoms moved freely about the
Ar core, showing large-scale movements akin to liquid-like behavior. The
intermediate, partially stacked configurations were also seen more
frequently. At the higher temperatures, the clusters often had enough energy
to overcome the barriers between the monolayer and stacked configuration
spaces, but because the stacked configuration subspace is so much smaller
than the monolayer configuration subspace, only a small fraction of the walk
was spent in the stacked subspace, resulting in few stacked configuration
quenches. But for temperatures near 7~K though, the movement
between the regions slowed down enough that the occasional forays into the
stacked configuration subspace resulted in the Metropolis walker being
trapped for long periods before escaping. Coexistence was not seen at lower
temperatures since the movements into the stacked subspace became so
infrequent that the walker remained essentially trapped in the monolayer
subspace.

This type of coexistence behavior between the stacked and monolayer
configurations is also similar to the behavior reported by Shelley, Le Roy
and Amar in their molecular dynamics study of SF$_6$-Ar$_9$
clusters.\cite{SLA} The authors observed that the cluster underwent two
different ``melting'' transitions, the first at $T \approx 15$~K marking the
onset of Ar atom mobility within the unimolecular layer surrounding the
SF$_6$ core, and the second at $T \approx 35$~K marking the onset of facile
Ar atom motion out of and back into this layer. Over a narrow intermediate
temperature range, the authors noted the coexistence between the
``liquid-like'' monolayer rearrangments and the solid-like large-amplitude
motion of the stacked two-layer configurations. One of the fundamental
differences between this coexistence and that seen for homogeneous Ar
clusters was that the transition from the liquid-like monolayer isomers to
the solid-like stacked isomers required an increase in energy; that is, the
``solid'' was obtained after heating the ``liquid.'' This spontaneous
reversing isomerization is analogous to the Ne$_{12}$Ar behavior, since the
large spikes in the Metropolis stacked isomer quench curves, as well as the
small, but nonzero, values afterward, occurred at higher temperatures ($T
\gtrsim 7$ K) than the temperature where the cluster first isomerized to the
monolayer form ($T = 6.4$~K), and had no quenches to the stacked
configuration at all.

The quench profiles for Ne$_9$Ar$_4$ and Ne$_{10}$Ar$_3$ show that much care
must be exercised in interpreting quench results. Both clusters have similar
profiles, differing mostly in the Ne$_9$Ar$_4$ curves showing abrupt changes
about 2~K sooner, but whereas the Ne$_{10}$Ar$_3$ quench profile was
consistent with the J-walking and Metropolis heat capacity results, the
Ne$_9$Ar$_4$ profile was not. Ne$_9$Ar$_4$ has a much larger number of
low-lying non-icosahedral isomers lying within $1\epsilon_{\mbox{\scriptsize
Ne-Ne}}$ of the lowest energy isomer than does Ne$_{10}$Ar$_3$, but the high
density of isomers above $1\epsilon_{\mbox{\scriptsize Ne-Ne}}$ from the
global minimum is similar in both clusters. The curve representing the sum of
the quenches to the higher energy isomers rises even more quickly in
Ne$_9$Ar$_4$ because of the contributions from the many isomers lying within
$1\epsilon_{\mbox{\scriptsize Ne-Ne}}$ of the ground state, suggesting that
the temperature region near 4~K be interpreted as a solid-liquid transition
region as well. The heat capacity results, though, were more akin to those
obtained for Ne$_{12}$Ar than to those obtained for Ne$_{10}$Ar$_3$ --- the
Ne$_9$Ar$_4$ heat capacity peak was also seen only in the J-walking results.
The Metropolis results showed a small rise in this region, but no discernible
peak, again implying that these higher energy isomers were not accessed
dynamically at these low temperatures, and so this region is not the
solid-liquid transition region. The Ne$_9$Ar$_4$ heat capacity curve has
another peak at 24.27~K having no counterpart in Ne$_{10}$Ar$_3$ that could
be identified as the solid-liquid transition peak, but it seems surprising
that replacing a single Ne atom with an Ar atom in Ne$_{10}$Ar$_3$ to form
Ne$_9$Ar$_4$ could shift the transition temperature from 7~K to 24~K.
Moreover, the peak is so broad that it is hard to imagine that it could
represent a solid-liquid transition spanning a range from 5 to 30~K.

Visually examining samples of the Ne$_9$Ar$_4$ configurations generated
during some of the Metropolis walks revealed that peak encompassed more than
just a solid-liquid transition. At 5~K, the cluster showed only solid-like
behavior for tens of thousands of passes, but at temperatures near 7~K, the
Ar atoms remained fixed as a solid-like tetrahedral sub-cluster undergoing
small amplitude oscillations, while a Ne atom occasionally moved from one
site to another. Although there are many low-lying non-icosahedral isomers
having similar energies, the barriers were too high for the Metropolis walks
to overcome at such low temperatures; the ergodic J-walking could cover such
large regions of configuration space, though, resulting in quenches to a wide
distribution of isomers, but with relatively few to any one. Thus the quench
profile appears similar to that of a cluster in the liquid region, even
though none of the atoms were undergoing the rapid large scale movements
characteristic of liquid-like behavior. As the Metropolis walk temperature
was increased, the Ne atoms underwent increasingly large amplitude motions,
eventually moving freely about the still solid-like Ar core. Much higher
temperatures were required before the Ar atoms also started undergoing large
amplitude motions and rearrangements from their tetrahedral sites. This
implies that Ne$_9$Ar$_4$ underwent differentiated melting, with the outer Ne
atoms ``melting'' well before the Ar core. At temperatures near the heat
capacity peak, the Ne$_9$Ar$_4$ clusters showed full liquid-like behavior,
but the Ne atoms would sometimes move quite far apart from one another, and
from the Ar atoms, which still tended to remain together, leading to
dissociated configurations that would collapse again into condensed
configurations after a short while.

The similar broad shapes of the heat capacity peaks for the other clusters
up to Ne$_3$Ar$_{10}$, as well as their similar peak temperatures, indicate
behavior similar to that of Ne$_9$Ar$_4$, with the peaks encompassing
``pre-melting'' isomerizations at low temperatures (evident where the
J-walking and Metropolis curves showed larger discrepancies), ``melting'' of
the outer Ne atoms (that is, free movement of the Ne atoms around an intact
Ar sub-cluster core), melting of the Ar-core to give a fully liquid-like
cluster (although with the Ne and Ar atoms more segregated than not), and
dissociation of the Ne atoms (which remained about the undissociated Ar atoms
because of the confining potential); full dissociation of the cluster was
evident in other peaks occurring at slightly higher temperatures than shown
in Fig.~\ref{Fig:Cv}. None of these ``stages'' had sharply defined
temperature ranges, but formed a continuum of behavior. As the number of Ar
atoms increased, though, these stages shifted to higher temperatures and
became increasingly differentiated, so that by Ne$_4$Ar$_9$, liquid-like
behavior dominated the heat capacity peak region.

The lowest energy isomers for the predominately Ar clusters Ne$_4$Ar$_9$ to
NeAr$_{12}$ were all Ar-core icosahedral-like structures having increasingly
larger gaps between them and their next highest energy isomers. This is
reflected in their quench profiles, shown in Fig.~\ref{Fig:Quench3}. The
temperature at which the fraction of quenches to the ground state isomers
began to drop increased as the number of Ar atoms increased. Ne$_2$Ar$_{11}$
has its two other Ar-core icosahedral-like permutational isomers having
similar energies at about $1\epsilon_{\mbox{\scriptsize Ne-Ne}}$ from the
ground state. Their quench curves show that they too began being accessed by
about 5~K, but while the second lowest energy isomer went on to contribute a
maximum of 20\% of the quenches at 18~K, the third isomer had very few
quenches throughout. This behavior is similar to that observed for Ar-Kr
clusters, where the different quench levels for icosahedral-like
permutational isomers was due primarily to simple combinatorics. As can be
seen in Fig.~\ref{Fig:Minima}, the third lowest-energy isomer has its two Ne
atoms occupying the axial positions of the icosahedron. The number of ways
that the two Ne and 11 Ar atoms can be combined to form this arrangement is
much smaller than the number of ways the other two permutational isomers can
be formed. The fourth lowest energy Ne$_2$Ar$_{11}$ isomer is a Ne-core
icosahedral-like isomer, and like most of the other Ne-core icosahedral-like
isomers for the other Ne-Ar clusters, it had very few quenches.

The quench profile for NeAr$_{12}$ is similar to that of Ar$_{13}$.
NeAr$_{12}$ has a Ne-core icosahedral-like isomer, as well as three truncated
Ar-core icosahedral-like isomers having the lone Ne atom located on one of
the icosahedral faces. The Ne-core icosahedral-like isomer again had very few
quenches, but the other three isomers began being accessed at about 20~K,
reaching a maximum at the heat capacity peak temperature. These three isomers
and the ground state isomer accounted for 75\% of the quenches at the heat
capacity peak temperature, which was similar to Ar$_{13}$, whose four lowest
isomers accounted for 80\% of the quenches at the heat capacity peak
temperature. The quench profile for NeAr$_{12}$ is very much different than
that of its ArKr$_{12}$ counterpart, though. Its analogous Kr-core
icosahedral-like isomer also dominated the quenches near the heat capacity
peak temperatures, but since this isomer was the higher energy one, it
started with no quenches at the very low temperatures where the ground state
Ar-core isomer was dominant. ArKr$_{12}$ had a moderately sized heat capacity
peak occurring below the solid-liquid transition peak that corresponded to
the transformation from the Ar-core to the Kr-core isomer. Because the
analogous Ne-core isomer was not the ground state isomer for NeAr$_{12}$,
there was no analogous transformation from it to the Ar-core isomer, and so
there was no smaller heat capacity peak for NeAr$_{12}$ corresponding to the
one for ArKr$_{12}$.

% Ar-Kr Cv paper: discussion section
\section{Conclusion}                                \label{Sec:conclusion}
The results obtained for 13-atom Ne-Ar clusters were very much different than
those obtained for their Ar-Kr counterparts. Where the Ar-Kr cluster
properties were mostly alike and similar to those of their homogeneous
counterpart, the Ne-Ar cluster properties exhibited a wide range of diverse
and unique behavior that showed little semblance to their corresponding
homogeneous cluster properties. For example, Ne$_{12}$Ar and Ne$_{11}$Ar$_2$
clusters underwent transitions from stacked Ne-core icosahedral-like
configurations to Ar-core monolayer configurations that none of the other
clusters did. Ne$_4$Ar$_9$ and Ne$_3$Ar$_{10}$ were much alike, but their
complements Ne$_9$Ar$_4$ and Ne$_{10}$Ar$_3$ were very different from one
another. All the Ar-Kr clusters showed the magic-number behavior
characteristic of Ar$_{13}$, but except for NeAr$_{12}$, none of the Ne-Ar
clusters did. Analogous to Ar$_{13}$, whose icosahedral and much higher
energy truncated icosahedral isomers dominated the solid-liquid transition
region, the Ar-Kr clusters' solid-liquid transition regions were dominated by
closely spaced Ar-core icosahedral-like permutational isomers and higher
energy truncated icosahedral-like permutational isomers. For the Ne-Ar
clusters, though, the corresponding Ne-core icosahedral-like and truncated
icosahedral-like permutational isomers played important roles only for
Ne$_{12}$Ar and Ne$_{11}$Ar$_2$, and were conspicuously absent for the
others. Instead, the low-lying Ne-Ar isomers were dominated by different
types of non-icosahedral isomers, many of which had no homogeneous
counterparts. The permutational isomers derived from the low-energy Ne-Ar
topological forms did not have energies lying in closely spaced groups that
were separated from one another by large gaps, as was the case for the Ar-Kr
clusters. Instead, the Ne-Ar permutational isomer energies for
the various topological forms were typically spread out over ranges that were
comparable or larger than the differences between the different sets, leading
to very interspersed isomer energy spectra; most of the Ne-Ar clusters were
devoid of any large gaps in their minima energy spectra.

These large differences in behavior between the 13-atom Ne-Ar and Ar-Kr
clusters result primarily from the disparateness between the Ne and Ar atomic
sizes compared to the much more similar Ar and Kr sizes, and to a lesser
extent, from the large difference between the Ne and Ar intermolecular
forces compared to the Ar and Kr forces. The former primarily effected the
cluster packing efficiency, decreasing the extraordinary stability of the
icosahedral structure to such an extent that the icosahedral-like isomers
energies became comparable to the non-icosahedral isomer energies for most of
the Ne-Ar clusters; this mostly effected the clusters' low-temperature
behavior. The latter was predominant at higher temperatures and resulted in
extended solid-liquid transition regions. Ar-Kr clusters, like their
homogeneous counterpart, underwent solid-liquid transitions over temperature
ranges that, while not sharply delimited, could be nonetheless differentiated
by the abrupt changes in many of their properties. The Ar and Kr atoms were
similar enough that the modifications in the clusters' transition behavior
were largely minor. For many of the Ne-Ar clusters, though, as the
temperature increased from the solid region, the clusters underwent a
pre-melting stage where the Ar atoms remained fixed as a solid sub-cluster
core undergoing low amplitude motions, while the Ne atoms wrapped around the
Ar core, slowly moving from site to site. Increasing the temperature led to
liquid-like motion of the Ne atoms as they moved with little hindrance around
the still solid-like Ar sub-cluster core. For those clusters having more than
4 Ar atoms, much higher temperatures were required before the Ar core also
showed liquid-like behavior. At these temperatures, the atoms moved freely
about one another, although the Ne and Ar atoms still tended to segregate,
with the Ar atoms mostly in the center. Similarly, as the temperature
increased from the liquid region, the Ne atoms dissociated first, resulting
in a liquid-like Ar-atom sub-cluster surrounded by freely moving Ne atoms,
confined within the constraining volume. These transitions each had extended
temperature regions of several degrees that mostly overlapped, so that the
overall solid-liquid transitions occurred over a much wider temperature
range, resulting in very broad heat capacity peaks with peak heights that
were mostly much lower than those obtained for the Ar-Kr clusters. As the
number of Ne atoms diminished, the melting behavior transformed into the more
usual differentiated behavior, so that by NeAr$_{12}$, the melting behavior
was quite similar to that of Ar$_{13}$.

The results obtained for the Ne-Ar clusters again demonstrated the need to
consider several cluster properties together, in context to one another, when
attempting to interpret their underlying behavior. While the J-walking
simulations were able to provide accurate heat capacities and quench
profiles, these were insufficient in themselves to interpret the solid-liquid
transition behavior. The J-walking results for Ne$_{12}$Ar showed a
pronounced heat capacity peak that seemed indicative of a solid-liquid
transition, and the quench profiles were consistent with such an
interpretation. However the Metropolis heat capacity and quench results, as
well as visual inspections of the configurations obtained during Metropolis
walks, indicated that the solid-liquid transition region occurred at higher
temperatures. The Ne$_{12}$Ar J-walking heat capacity peak was due to the
increased fluctuations in the cluster's energies arising from the ergodic
sampling of the stacked isomer's and monolayer isomers' configuration
subspaces --- the peak occurred at too low a temperature for dynamic motions
to overcome the barriers between these regions, and so it was more akin to
the low temperature mixing anomaly peaks observed in the Ar-Kr clusters than
to their solid-liquid transition peaks. Similarly, the J-walker quench
profiles for Ne$_{10}$Ar$_3$ and Ne$_9$Ar$_4$ were alike, while their heat
capacity curves and underlying solid-liquid transition behavior were quite
different. Since I performed no molecular dynamics simulations on these
clusters, I have no direct dynamical evidence of the coexistence behavior
that is an essential characteristic of homogeneous 13-atom rare gas clusters,
but the J-walking and Metropolis results can be used to infer changes in the
coexistence behavior for the different cluster compositions. NeAr$_{12}$ and
possibly Ne$_{12}$Ar had sufficient implicit evidence to infer they underwent
coexistence, although the coexistence behavior in Ne$_{12}$Ar was different
from that of Ne$_{13}$. For the other clusters, though, the solid-liquid
transitions were best described as a progressive evolution of isomerizations
between predominately topologically distinct isomers. This progression became
less continuous as the proportion of Ar atoms increased and the different
sets of permutational isomers became increasingly segregated, eventually
transforming into the coexistence behavior of NeAr$_{12}$ and Ar$_{13}$. It
is interesting to note that the broadening and lowering of the heat capacity
peaks as the cluster composition varied from NeAr$_{12}$ to Ne$_5$Ar$_8$,
followed by the narrowing and increased peak heights from Ne$_6$Ar$_7$ to
Ne$_9$Ar$_4$, were very similar to the changes in the homogeneous cluster
heat capacities as the aggregate size $N$ changed from 13 to
19.\cite{Magic_Cv} The large $N = 13$ peak decreased and broadened as $N$
increased to 16, and then narrowed and increased to the $N = 19$ peak. The
magic number endpoints $N = 13$ and 19 had pronounced solid-liquid
coexistence ranges, while the clusters $N = 15, 16$ and 17, which had low,
broad peaks, underwent solid-liquid transitions comprised of smooth
progressions of isomerizations.

As was the case with the Ar-Kr clusters, the calculations presented here have
demonstrated that the standard Metropolis algorithm is inadequate for dealing
with the low-temperature behavior of heterogeneous clusters. Unlike the Ar-Kr
case, though, augmenting the Metropolis method with the atom-exchange method
did not improve its accuracy much for the Ne-Ar clusters. Again, the
J-walking calculations for these binary clusters were more complicated than
those done on similar homogeneous clusters because of the greater possibility
for systematic errors corrupting the J-walker distributions. Using
multiple trials from independently generated distributions helped ensure that
such errors were not present. The computational overhead was
mitigated by the shorter walk lengths required for J-walking to achieve the
desired level of accuracy compared to the Metropolis method; for example, for
the Ne$_{11}$Ar$_2$ heat capacity curve over the range $6 \leq T \leq 8$ K,
the J-walker runs of $10^6$ passes each had a level of accuracy that required
runs totaling $10^8$ passes for the Metropolis method to match, and even at
$10^8$ total passes, the Metropolis method had large systematic errors below
6~K.

Finally, it should be noted that the Ne-Ar results reported here are
deficient in that they were obtained from classical simulations. Previous
quantum studies of Ne clusters have shown that quantum effects play a large
role in their behavior,\cite{JWFPI} and so quantum simulations would likely
be needed for a realistic treatment of Ne-Ar clusters as well, especially for
the predominately Ne ones. Since my primary motivation with this study was to
elucidate the effects of largely different atomic sizes and intermolecular
forces on binary cluster behavior, as well as the dependence on cluster
composition, the Ne-Ar clusters served primarily as a model system, and so
the classical simulations were quite adequate. Quantum effects are most
pronounced at lower temperatures, and for the predominately Ne clusters, much
of the interesting behavior occurred at the low temperatures where quantum
effects are expected to have a major impact. For example, barriers between
low-lying isomers are effectively reduced in quantum simulations, and so
transitions from stacked isomers to monolayer isomers in Ne$_{12}$Ar could
become dynamically accessible in quantum simulations, and the heat capacity
peak behavior thus much modified.

\acknowledgments
The support of the Natural Sciences and Engineering Research Council of
Canada (NSERC) for this research is gratefully acknowledged. I thank the
University of Lethbridge Computing Services for generously providing me the
use of their workstations for some of the calculations reported, and the
University of Waterloo for the use of their facilities. I also thank David
L. Freeman for his helpful discussions.

% Ar-Kr Cv paper: references

% Ar-Kr Cv paper: figure captions
\begin{figure}
\caption{Heat capacity curves for Ne$_{11}$Ar$_2$ clusters. The solid
curves were obtained from separate J-walking runs, each using a different set
of J-walker distributions; the data for each temperature were obtained from
$10^6$ total passes. Over most of the temperature domain, the differences
between the curves are comparable to the noise levels inherent in each,
indicating that the systematic errors associated with each J-walker
distribution were sufficiently small in these regions. The larger
differences between the curves at the low-temperature peak illustrate how
sensitive the peak is to small systematic errors in the low-temperature
J-walker distributions. Also included is the dashed curve obtained from
standard Metropolis runs having $10^7$ total passes of data accumulation at
each temperature. The curve was begun at the lowest temperature using the
lowest energy isomer as the initial configuration, with the final
configuration at each temperature then used as the initial configuration for
the next temperature. Substantial discrepancies due to quasiergodicity are
evident for temperatures below 6~K, with the low-temperature peak completely
absent. Augmenting standard Metropolis with the atom-exchange algorithm did
not improve its low temperature performance in this case, as can be seen by
the filled circles, which lie mostly on the Metropolis curve. The open
circles represent results obtained from standard Metropolis runs having
$10^8$ total passes of data accumulation at each temperature. These are in
better agreement with the J-walker results down to 6~K, but still show the
same sharp drop at this temperature that the shorter Metropolis runs did.
The heat capacities are in reduced units.
\label{Fig:multi_trials}}
\end{figure}

\begin{figure}
\caption{The 13 lowest energy isomers found for Ne$_{13-n}$Ar$_n$ clusters
($1 \leq n \leq 12$), in order of increasing energy; their energies are
listed in Table~\protect\ref{Tbl:Minima}. The isomers were obtained from
quenches of the configurations stored in J-walker distribution files. For the
predominately-Ne clusters Ne$_{12}$Ar and Ne$_{11}$Ar$_2$, the global minimum
corresponds to an icosahedral-like configuration having a Ne atom as the
central atom, but for the predominately-Ar clusters Ne$_4$Ar$_9$ to
NeAr$_{12}$, the global minimum in each case corresponds to an
icosahedral-like configuration having an Ar atom as the central atom. The
global minimum for each of the intermediate cases Ne$_{10}$Ar$_3$ to
Ne$_5$Ar$_8$ is non-icosahedral.
\label{Fig:Minima}}
\end{figure}

\begin{figure}
\caption{Potential energies as functions of composition for the lowest energy
non-icosahedral isomers (circles) and icosahedral-like isomers (squares for
Ne-core, and diamonds for Ar-core isomers). The energies in the upper plot
are in reduced units of $\epsilon_{\mbox{\scriptsize Ne-Ne}}$, while the
energies in the lower plot have been scaled by the composition weighted
average $\epsilon_{\rm Avg} = X_{\rm Ne} \epsilon_{\mbox{\scriptsize Ne-Ne}}
+ X_{\rm Ar} \epsilon_{\mbox{\scriptsize Ar-Ar}}$, where $X_{\rm Ne} = n_{\rm
Ne}/(n_{\rm Ne} + n_{\rm Ar})$ and $X_{\rm Ar} = 1 - X_{\rm Ne}$. The Ne-core
icosahedral-like isomers were the global minimum configurations up to
Ne$_{11}$Ar$_2$, after which the non-icosahedral isomers became the ground
state. The Ar-core icosahedral-like configurations became the ground
state at Ne$_4$Ar$_9$; for clusters having fewer than 7 Ar atoms, the Ar-core
icosahedral-like isomers were highly distorted, and those having fewer than 4
Ar atoms were unstable.
\label{Fig:MinimaComp}}
\end{figure}

\begin{figure}
\caption{Local potential energy minima (in reduced units) for
Ne$_{13-n}$Ar$_n$ clusters ($0 \leq n \leq 13$). In each case, the zero
energy level corresponds to the global minimum energy. The energies were
obtained from quenches of the configurations stored in J-walker distribution
files.
\label{Fig:Min_Energies}}
\end{figure}

\begin{figure}
\caption{Heat capacities for the binary clusters Ne$_{13-n}$Ar$_n$ ($1 \leq n
\leq 12$). In each case, the thick curve represents the J-walking results and
the thin curve the standard Metropolis results; the dots represent the
results obtained using Metropolis Monte Carlo augmented with the
atom-exchange algorithm. Large discrepancies between the Metropolis and
J-walking results can be seen to occur at lower temperatures for all the
clusters except for NeAr$_{12}$. The atom-exchange method showed some
improvement over the standard Metropolis method, but still had large
deviations from the J-walking results. For the predominately Ar clusters
Ne$_3$Ar$_{10}$ to NeAr$_{12}$, the large high-temperature peaks correspond
to solid-liquid transitions, but for the other clusters, the peaks are so
broad that they encompass not only the solid-liquid transition, but also the
dissociation of the Ne component at the higher temperatures. The Ne$_{12}$Ar
peak and the very small low-temperature peaks seen in Ne$_{11}$Ar$_2$,
Ne$_9$Ar$_4$, Ne$_8$Ar$_5$, Ne$_6$Ar$_7$ and Ne$_5$Ar$_8$ are due to
low-lying isomers whose energies are close to the global minimum energies.
Only NeAr$_{12}$ has a heat capacity curve similar to that of its homogeneous
counterpart.
\label{Fig:Cv}}
\end{figure}

\begin{figure}
\caption{Trends in the higher temperature heat capacity peak parameters. The
top plot shows the peak temperatures as a function of cluster composition,
while the middle plot shows the peak heights as a function of composition.
The bottom plot depicts the peak heights as a function of the peak
temperature. The uncertainties in the peak heights and temperatures are
smaller than the symbol size. The predominately Ne clusters ($n_{\rm Ne} >
9$) and the predominately Ar clusters ($n_{\rm Ar} > 9$) have parameter
trends much different than that of the other clusters. The peak temperature
and height for each cluster also are listed in
Table~\protect\ref{Tbl:Cv_peaks}.
\label{Fig:Cv_Peaks}}
\end{figure}

\begin{figure}
\caption{Potential energy curves as functions of temperature for
Ne$_{13-n}$Ar$_n$ clusters ($0 \leq n \leq 13$). The curves were obtained
from the J-walking simulations.
\label{Fig:Caloric}}
\end{figure}

\begin{figure}
\caption{Quench results for the predominately Ne clusters. These were obtained
by periodically quenching cluster configurations by steepest descent every
1000 passes during one of the J-walker runs, giving 1000 quenched
configurations for each temperature. In each case, results for the six
lowest energy isomers shown in Fig.~\protect\ref{Fig:Minima} are plotted,
and the solid curve represents the sum of all the other isomers. The dashed
vertical lines in each plot indicate the heat capacity peak temperatures.
\label{Fig:Quench1}}
\end{figure}

\begin{figure}
\caption{J-walker quench results for the clusters Ne$_8$Ar$_5$ to
Ne$_5$Ar$_8$.
\label{Fig:Quench2}}
\end{figure}

\begin{figure}
\caption{J-walker quench results for the predominately Ar clusters. The
NeAr$_{12}$ curves are similar to those of Ar$_{13}$.
\label{Fig:Quench3}}
\end{figure}

\begin{figure}
\caption{Metropolis quench results for Ne$_{12}$Ar clusters. The solid
symbols represent the results obtained from Metropolis simulations that were
started from the global minimum configuration at 0.1~K, with the temperature
increased in increments of 0.1~K, while the open symbols represent the
results obtained from similar runs started at 10~K, with the temperature
decreased in increments of 0.1~K. The sets of curves agree with each other,
and agree with the J-walking curves for Ne$_{12}$Ar shown in
Fig.~\protect\ref{Fig:Quench1} only for temperatures greater than 6.2~K. The
Metropolis simulations run in the direction of decreasing temperature did not
reach the lowest energy isomer as the temperature approached zero.
\label{Fig:QuenchMet}}

\end{figure}

% Ne-Ar Cv paper: table 1
\begin{table}
\caption{Lennard-Jones parameters used in the calculations. The Ne-Ne and
Ar-Ar parameters were obtained from Ref.~\protect\onlinecite{LDW}. The Ne-Ar
parameters were obtained from the usual Lorentz-Berthelot mixing rules, with
$\epsilon_{\mbox{\scriptsize Ne-Ar}} = (\epsilon_{\mbox{\scriptsize
Ne-Ne}}\,\epsilon_{\mbox{\scriptsize Ar-Ar}})^{1/2}$ and
$\sigma_{\mbox{\scriptsize Ne-Ar}} = \frac{1}{2}(\sigma_{\mbox{\scriptsize
Ne-Ne}} + \sigma_{\mbox{\scriptsize Ar-Ar}}).$}
\begin{tabular}{lrrr}
\multicolumn{1}{c}{Parameter} & \multicolumn{1}{c}{Ne-Ne} &
\multicolumn{1}{c}{Ne-Ar} &
\multicolumn{1}{c}{Ar-Ar} \\
\hline
$\epsilon$/K & 35.60 & 65.20 & 119.4 \\
$\sigma/$\mbox{\AA} & 2.749 & 3.077 & 3.405 \\
\end{tabular}
\label{Tbl:LJ-params}
\end{table}

% Ne-Ar Cv paper: table2
\widetext
\begin{table}
\caption{Potential energies for the thirteen lowest-energy equilibrium
configurations shown in Fig.~\protect\ref{Fig:Minima}. These values were
obtained from quench studies of the configurations stored in the J-walker
distribution files or by minimizing the various permutations of icosahedral
configurations. The energies are expressed in reduced units of
$-\epsilon_{\mbox{\scriptsize Ne-Ne}}$.
\label{Tbl:Minima}}
\begin{tabular}{*{8}{r}}
\multicolumn{1}{c}{Isomer} &
\multicolumn{1}{c}{Ne$_{13}$} &
\multicolumn{1}{c}{Ne$_{12}$Ar} &
\multicolumn{1}{c}{Ne$_{11}$Ar$_{2}$} &
\multicolumn{1}{c}{Ne$_{10}$Ar$_{3}$} &
\multicolumn{1}{c}{Ne$_{9}$Ar$_{4}$} &
\multicolumn{1}{c}{Ne$_{8}$Ar$_{5}$} &
\multicolumn{1}{c}{Ne$_{7}$Ar$_{6}$} \\
\hline
 0 & 44.32680 & 50.14737 & 56.57433 & 64.24192 & 71.38399 & 78.86406 & 86.39936 \\
 1 & 41.47198 & 48.83695 & 56.52063 & 63.72413 & 71.08382 & 78.66373 & 85.83742 \\
 2 & 41.44460 & 48.76841 & 56.06123 & 63.71574 & 70.94101 & 78.52868 & 85.81146 \\
 3 & 41.39440 & 48.76223 & 56.03686 & 63.66647 & 70.93143 & 78.51267 & 85.76317 \\
 4 & 40.75851 & 48.71912 & 55.93749 & 63.59504 & 70.85554 & 78.46705 & 85.63794 \\
 5 & 40.72846 & 48.65682 & 55.81092 & 63.59501 & 70.77891 & 78.04509 & 85.62603 \\
 6 & 40.71041 & 48.38159 & 55.79632 & 63.19284 & 70.75207 & 78.02281 & 85.56162 \\
 7 & 40.67380 & 48.32417 & 55.70070 & 63.10945 & 70.74139 & 77.99609 & 85.55390 \\
 8 & 40.67017 & 48.31641 & 55.69579 & 63.05431 & 70.72857 & 77.93368 & 85.50783 \\
 9 & 40.61547 & 48.24531 & 55.64982 & 63.04676 & 70.72729 & 77.88402 & 85.49314 \\
10 & 40.60458 & 48.18147 & 55.61569 & 63.01544 & 70.68951 & 77.88031 & 85.41050 \\
11 & 40.54129 & 48.11949 & 55.61427 & 62.96995 & 70.67538 & 77.84458 & 85.39834 \\
12 & 40.43333 & 48.11635 & 55.60919 & 62.95865 & 70.65142 & 77.73255 & 85.36450 \\
\hline
\multicolumn{1}{c}{Isomer} &
\multicolumn{1}{c}{Ar$_{13}$} &
\multicolumn{1}{c}{NeAr$_{12}$} &
\multicolumn{1}{c}{Ne$_{2}$Ar$_{11}$} &
\multicolumn{1}{c}{Ne$_{3}$Ar$_{10}$} &
\multicolumn{1}{c}{Ne$_{4}$Ar$_{9}$} &
\multicolumn{1}{c}{Ne$_{5}$Ar$_{8}$} &
\multicolumn{1}{c}{Ne$_{6}$Ar$_{7}$} \\
\hline
 0 & 148.66910 & 137.68293 & 127.58029 & 118.43464 & 109.36661 & 100.77911 & 93.73315 \\
 1 & 139.09422 & 133.68855 & 126.78465 & 117.49640 & 108.72718 & 100.67782 & 93.61382 \\
 2 & 139.00238 & 133.57037 & 126.72343 & 116.79154 & 108.59267 & 100.67727 & 93.53146 \\
 3 & 138.83402 & 133.53295 & 124.53811 & 116.72628 & 108.52706 & 100.65515 & 93.47289 \\
 4 & 136.70131 & 133.46282 & 124.44729 & 116.54890 & 108.46116 & 100.63225 & 93.41254 \\
 5 & 136.60051 & 130.36349 & 124.43270 & 116.51740 & 108.45875 & 100.62103 & 93.41221 \\
 6 & 136.53996 & 129.95629 & 124.41134 & 116.50550 & 108.44127 & 100.56415 & 93.38494 \\
 7 & 136.41718 & 129.92105 & 124.38340 & 116.46722 & 108.41135 & 100.56237 & 93.33902 \\
 8 & 136.40501 & 129.86693 & 124.34131 & 116.44073 & 108.40618 & 100.56091 & 93.10541 \\
 9 & 136.22154 & 129.80847 & 124.34068 & 116.42079 & 108.35227 & 100.55547 & 93.07193 \\
10 & 136.18504 & 129.70576 & 124.32050 & 116.20818 & 108.25224 & 100.51264 & 92.99134 \\
11 & 135.97277 & 129.70212 & 123.66593 & 115.98123 & 107.99351 & 100.50910 & 92.94263 \\
12 & 135.61066 & 129.69684 & 123.65838 & 115.74204 & 107.90823 & 100.47166 & 92.87926 \\
\end{tabular}
\end{table}
\newpage

% Ne-Ar Cv paper: table3
\widetext
\begin{table}
\caption{Heat capacity peak parameters for 13-atom Ne-Ar clusters. These
values were obtained by smoothing and interpolating the J-walking data shown
in Fig.~\protect\ref{Fig:Cv}. The Ne$_{13}$ and Ar$_{13}$ values were
obtained by scaling the values in reported Ref.~\protect\onlinecite{ArKr_Cv}.
The heat capacity uncertainty estimates are averages of single standard
deviations of the points near the peaks.
\label{Tbl:Cv_peaks}}
%\squeezetable
\begin{tabular}{ldddd}
  &
\multicolumn{2}{c}{Lower temperature peak} &
\multicolumn{2}{c}{Higher temperature peak} \\
\cline{2-3}
\cline{4-5}
  &
\multicolumn{1}{c}{$T_{\rm peak}$ (K)} &
\multicolumn{1}{c}{$\langle C_V^*\rangle_{\rm peak}$} &
\multicolumn{1}{c}{$T_{\rm peak}$ (K)} &
\multicolumn{1}{c}{$\langle C_V^*\rangle_{\rm peak}$} \\
\hline
Ne$_{13}$       &                 &
                & 10.15 $\pm$0.02 & 118.5 $\pm$0.6 \\
Ne$_{12}$Ar     &                 &
                &  5.20 $\pm$0.03 &  67.1 $\pm$0.9 \\
Ne$_{11}$Ar$_2$ &  0.37 $\pm$0.03 &  41.1 $\pm$0.7
                &  8.92 $\pm$0.14 &  59.2 $\pm$0.3 \\
Ne$_{10}$Ar$_3$ &                 &
                &  6.76 $\pm$0.03 &  58.5 $\pm$0.5 \\
Ne$_9$Ar$_4$    &  3.40 $\pm$0.02 &  47.4 $\pm$0.7
                & 24.27 $\pm$0.34 &  79.6 $\pm$0.3 \\
Ne$_8$Ar$_5$    &  2.02 $\pm$0.02 &  40.6 $\pm$0.3
                & 24.46 $\pm$0.17 &  75.4 $\pm$0.4 \\
Ne$_7$Ar$_6$    &                 &
                & 25.62 $\pm$0.20 &  71.7 $\pm$0.4 \\
Ne$_6$Ar$_7$    &  1.69 $\pm$0.06 &  38.9 $\pm$0.3
                & 24.83 $\pm$0.35 &  69.3 $\pm$0.4 \\
Ne$_5$Ar$_8$    &  1.63 $\pm$0.10 &  39.8 $\pm$0.4
                & 25.40 $\pm$0.33 &  67.3 $\pm$0.3 \\
Ne$_4$Ar$_9$    &                 &
                & 22.42 $\pm$0.05 &  68.2 $\pm$0.5 \\
Ne$_3$Ar$_{10}$ &                 &
                & 22.01 $\pm$0.31 &  73.9 $\pm$0.4 \\
Ne$_2$Ar$_{11}$ &                 &
                & 25.49 $\pm$0.14 &  84.2 $\pm$0.7 \\
NeAr$_{12}$     &                 &
                & 29.68 $\pm$0.11 &  95.2 $\pm$0.5 \\
Ar$_{13}$       &                 &
                & 34.05 $\pm$0.07 & 118.0 $\pm$0.6 \\
\end{tabular}
\end{table}


\begin{references}
\bibitem{Sav_Gol} The Savitsky-Golay algorithm was used for the data
    smoothing, and 2D cubic splines were used for the interpolation. These
    routines are implemented in the MS Windows graphics software package
    Axum, version 4.0.
\bibitem{DissPeak} The Metropolis simulations were extended to much higher
    temperatures than the J-walking simulations. Although not shown in the
    plots in Fig.~\ref{Fig:Cv}, the Metropolis curves for each cluster had
    a higher temperature peak asscociated with the cluster's dissociation
    region. These peaks' locations and heights depended very much on the
    boundary conditions imposed by the constraining spheres.
\end{references}
\end{document}